\documentclass[fleqn,usenatbib]{mnras}

\usepackage{newtxtext,newtxmath}
\usepackage[T1]{fontenc}
\usepackage{ae,aecompl}
\usepackage{float}
\usepackage{eso-pic}

\newcommand{\plasmabeta}{\beta_{\mathrm{mag}}}
\newcommand{\emiss}{\beta_{\mathrm{em}}}

\newcommand{\mach}{\mathcal{M}}

\newcommand{\sfrunits}{\msol\,\mathrm{yr}^{-1}\,\mathrm{kpc}^{-2}}

\newcommand{\lcob}{L'_\mathrm{{C\,O\,(5-4)}}}
\newcommand{\lcoa}{L'_\mathrm{{C\,O\,(1-0)}}}
\newcommand{\Clump}{\mathrm{Clump\,A}}
\newcommand{\clump}{\mathrm{clump\,A}}

\newcommand{\msol}{\mbox{$\mathrm{M}_{\sun}$}}
\newcommand{\lsol}{\mbox{$\mathrm{L}_{\sun}$}}

\newcommand{\tff}{t_\mathrm{ff}}

\newcommand{\pc}{\mathrm{pc}}
\newcommand{\kpc}{\mathrm{kpc}}

\newcommand{\yr}{\mathrm{yr}}

\newcommand{\sigsfr}{\Sigma_\mathrm{SFR}}
\newcommand{\siggas}{\Sigma_\mathrm{gas}}

\newcommand{\sigmasfr}{\Sigma_{\mathrm{SFR}}}

\usepackage{etoolbox}
\makeatletter
\patchcmd\@combinedblfloats{\box\@outputbox}{\unvbox\@outputbox}{}{%
   \errmessage{\noexpand\@combinedblfloats could not be patched}%
}%
 \makeatother

\usepackage{graphicx}	
\usepackage{amsmath}	
\usepackage{amssymb}	

\title[IV Submission Clean Testing Star Formation Laws on SDP 81 Using ALMA]{Testing Star Formation Laws in a Starburst Galaxy At Redshift 3 Resolved with ALMA}

\author[P. Sharda et al.]{
P. Sharda$^{2,3,4}$\thanks{E-mail: f2013440@pilani.bits-pilani.ac.in (PS)},
C. Federrath$^{1}$,
E. da Cunha$^{1}$, 
A. M. Swinbank$^{5,6}$ and
S. Dye$^{7}$
\\
$^{1}$Research School of Astronomy and Astrophysics, The Australian National University, Canberra, ACT~2611, Australia\\
$^{2}$Department of Physics, Birla Institute of Technology and Science, Pilani, Rajasthan 333031, India\\
$^{3}$Department of Electrical and Electronics Engineering, Birla Institute of Technology and Science, Pilani, Rajasthan 333031, India\\
$^{4}$Department of Physics and Astronomy, University of Exeter, Stoker Road, Exeter EX4 4QL, UK\\
$^{5}$Institute for Computational Cosmology, Durham University, Durham DH1 3LE, UK\\
$^{6}$Centre for Extragalactic Astronomy, Department of Physics,
Durham University, Durham, DH1 3LE, UK\\
$^{7}$School of Physics and Astronomy, University of Nottingham, University Park, Nottingham NG7 2RD, UK
}

\date{Accepted 2018 March 26. Received 2018 March 26; in original form 2017 December 11}

\pubyear{2018}
\begin{document}
\label{firstpage}
\pagerange{\pageref{firstpage}--\pageref{lastpage}}
\maketitle

\begin{abstract}
Using high-resolution (sub-kiloparsec scale) data obtained by ALMA, we analyze the star formation rate (SFR), gas content and kinematics in SDP 81, a gravitationally-lensed starburst galaxy at redshift $3$. We estimate the SFR surface density ($\sigmasfr$) in the brightest clump of this galaxy to be $357^{+135}_{-85}\,\sfrunits$, over an area of $0.07\pm0.02\,\kpc^2$. Using the intensity-weighted velocity of CO$\,$(5-4), we measure the turbulent velocity dispersion in the plane-of-the-sky and find $\sigma_{\mathrm{v,turb}} = 37\pm5\,\mathrm{km\,s}^{-1}$ for the clump, in good agreement with previous estimates along the line of sight, corrected for beam smearing. Our measurements of gas surface density, freefall time and turbulent Mach number allow us to compare the theoretical SFR from various star formation models with that observed, revealing that the role of turbulence is crucial to explaining the observed SFR in this clump. While the Kennicutt Schmidt (KS) relation predicts an SFR surface density of $\Sigma_{\mathrm{SFR,KS}} = 52\pm17\,\sfrunits$, the single-freefall model by Krumholz, Dekel and McKee (KDM) predicts $\Sigma_{\mathrm{SFR,KDM}} = 106\pm37\,\sfrunits$. In contrast, the multi-freefall (turbulence) model by Salim, Federrath and Kewley (SFK) gives $\Sigma_{\mathrm{SFR,SFK}} = 491^{+139                                                         }_{-194}\,\sfrunits$. Although the SFK relation overestimates the SFR in this clump (possibly due to the negligence of magnetic fields), it provides the best prediction among the available models. Finally, we compare the star formation and gas properties of this galaxy to local star-forming regions and find that the SFK relation provides the best estimates of SFR in both local and high-redshift galaxies.
\end{abstract}



\section{Introduction}
\label{s:intro}

Numerous star formation relations have been proposed in a quest to universalize the theory of star formation, by linking the star formation rate (SFR) with the mass of gas, its freefall time, virial parameter, magnetic field strength and turbulence (\citealt{1997ApJ...481..703S,1998ARA&A..36..189K,2002ApJ...577..206E,2011ApJ...733...87S,2012ApJ...745...69K} (hereafter, KDM12); \citealt{2012ApJ...760L..16R,2015ApJ...804...54E,2015ApJ...805..145E,2015ApJ...806L..36S} (hereafter, SFK15); \citealt{2015ApJ...814L..30E,2016ApJ...833...23N,2017A&A...602L...9M}). While these relations have been shown to be valid for star-forming regions in the Milky Way and local galaxies \citep{2008AJ....136.2846B}, the lack of spatial resolution has limited us in testing them on high-redshift sources with $z>1$. Thanks to the high spatial resolution of ALMA, several high-redshift galaxies emitting in the submillimeter (sub-mm) regime have been detected and resolved over the last few years \citep{2016ApJ...833...70D,2016ApJ...826..112S,2017ApJ...842L..16L,2017ApJ...840...78D,2017A&A...608A..15B}, particularly if they are gravitationally-lensed by a foreground source \citep{1997ApJ...490L...5S,2002MNRAS.331..495S,2013ApJ...767..132H,2017ApJ...843L..21J,2017ApJ...836L...2B,2017ApJ...837L..21L,2017ApJ...843L..35W,2017MNRAS.472.2028F}. These galaxies are known to be rigorous sites of dusty star formation where molecular gas plays a key role in modifying the structure of clusters where star formation occurs. Tracing molecular gas in these regions can give us valuable insight on the star formation characteristics of these galaxies since it is now known that molecular gas has a strong correlation with SFR whereas atomic gas does not \citep{2002ApJ...569..157W,2008AJ....136.2846B,2009ApJ...704..842B}.

J090311.6+003906 (hereafter, referred to as SDP 81) was detected as a lensed galaxy in the $H-ATLAS$ survey of bright submillimeter galaxies (SMGs) by \cite{2010Sci...330..800N} where the redshift was measured as $z = 3.042\pm0.001$ through ground based CO measurements. It falls in the popular definition of SMGs where the $850\,\mu\mathrm{m}$ flux density $S_{850} > 3\,\mathrm{mJy}$ and infrared luminosity $L_{\mathrm{IR}} \gtrsim 10^{12}\,\lsol$ \citep{2006ApJ...650..592K,2008MNRAS.384.1597C,2011ApJ...743..159H}. SDP 81 has also been established to be a dusty star-forming galaxy in previous works (\citealt{2014MNRAS.440.1999N}; \citealt{2015ApJ...806L..17S} (hereafter, S15); \citealt{2015MNRAS.452.2258D,2015PASJ...67...93H,2015MNRAS.451L..40R,2015MNRAS.453L..26R,2015PASJ...67...72T,2015ApJ...811..115W,2016ApJ...823...37H,2016MNRAS.457.2936I}). Even though there are significant uncertainties in determining the stellar mass of SDP 81, we note that it lies 1-2 orders of magnitude above the main sequence on the stellar mass $-$ star formation rate ($\mathrm{M}_{\star}-\mathrm{SFR}$) plane \citep{2014ApJS..214...15S,2015A&A...575A..74S,2015ApJS..219....8C,2015ApJ...808L..49G}. Thus, SDP 81 falls under the category of extreme starburst galaxies and is an ideal candidate to test star formation relations. This is further confirmed by the position of SDP 81 on the star formation rate $-$ gas mass ($\mathrm{SFR-M_{gas}}$) plane \citep{2014ApJ...793...19S}.

Our goal in this work is to extract the SFR in individual \textit{clumps} of this galaxy and compare it with that predicted by existing star formation relations. We refer to clumps as giant star-forming regions \citep{2006Natur.442..786G,2009ApJ...701..306E,2014ApJ...780...57B} substantially more massive and star-forming than typical molecular clouds in the Milky Way \citep{1995AJ....110.1576C,1996AJ....112..359V,2010MNRAS.403L..36S}, and possibly showing high star formation efficiencies \citep{2013A&A...553A.130F,2015Natur.521...54Z,2017MNRAS.469.4683C}. The paper is organized as follows: in Section \ref{s:data}, we summarize data reduction through lens modeling to create source plane reconstructed images \citep{2015MNRAS.452.2258D}. This section also identifies different clumps in this galaxy extracted by S15. Section \ref{s:dustsed} follows the dust spectral energy distribution (SED) fitting of a modified blackbody (MBB) and estimation of SFR surface density in the galaxy. We describe the kinematic analysis of CO$\,$(5-4) used to estimate the Mach number in Section \ref{s:kinematics}. In Section \ref{s:gasmass}, we present our estimates of the local gas mass and freefall time in the galaxy. Finally, we put all these parameters together to test various star formation relations and compare with the SFR surface density deduced through dust SED fitting in Section \ref{s:compare}. We summarize our findings in Section \ref{s:summ}.

We adopt the $\Lambda$CDM cosmology with H$_0 = 72\,\mathrm{km\,s}^{-1}\,\mathrm{Mpc}^{-1}$, $\Omega_\mathrm{m}$ = 0.27, $\Omega_{\Lambda}$ = 1-$\Omega_\mathrm{m}$ and a Chabrier IMF \citep{2003ApJ...586L.133C}. The luminosity distance and scale length corresponding to these parameters is $25.9\,\mathrm{Gpc}$ and $7.69\,\kpc\,\mathrm{arcsec}^{-1}$, respectively, for $z \approx 3.042$ \citep{2006PASP..118.1711W}.

\section{Data Reduction and Analysis}
\label{s:data}
ALMA observations of SDP 81 (RA = $09^\mathrm{h}03^\mathrm{m}11.57^\mathrm{s}$, Dec = $+00^{\circ}39'06.6''$) were taken during Science Verification cycle in October 2014. In the calibrated data\footnote{\url{https://almascience.nao.ac.jp/alma-data/science-verification}}, the lensed galaxy is seen in the form of an Einstein ring, with two arcs on the eastern and western sides \citep{2014MNRAS.440.2013D,2015ApJ...808L...4A}. Through $u\nu$ tapering, a resolution of $\sim150\times120\,\mathrm{mas}$ was achieved in the three bands (see Tables 1 and 3 of \cite{2015ApJ...808L...4A} for observed fluxes and noise levels). The CO$\,$(5-4) velocity cubes were binned to a velocity resolution of $21\,\mathrm{km}\,\mathrm{s}^{-1}$ \citep{2015ApJ...808L...4A}.

We use the source plane reconstructed images of continuum emissions (in ALMA Bands 4, 6 and 7, corresponding to $\lambda_{\mathrm{obs}}=2.0,1.3$ and $1.0$ mm, respectively) and CO$\,$(5-4) flux and velocity, created by S15, using the lensing model by \cite{2015MNRAS.452.2258D}. This model was used in the image plane with the semi-linear inversion method \citep{2003ApJ...590..673W} worked upon by \cite{2015MNRAS.452.2940N}. The average luminosity weighted magnification factors derived by \cite{2015MNRAS.452.2258D} for the continuum in band 6 and 7 are $15.8 \pm 0.7$ and $16.0 \pm 0.7$ respectively. This magnification is representative of a higher resolution by a factor of $\sim30$ (sub-kpc scale) than that in the typical non-lensed case \citep{2015ApJ...810..133I,2015ApJ...799...81S}.

\begin{figure}
\includegraphics[width=1.0\linewidth]{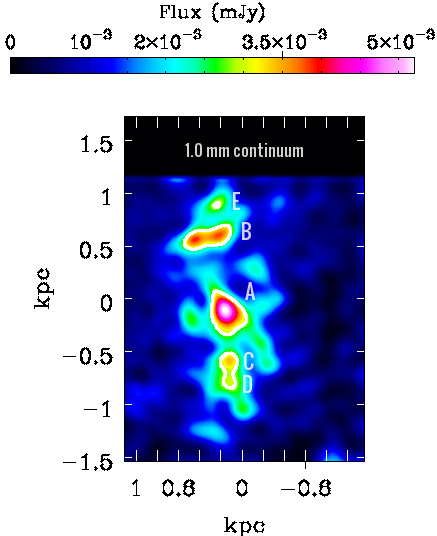}
\caption{Source reconstructed image of continuum flux in ALMA band 7 ($\lambda_{\textnormal{obs}}$ = 1.0 mm), created using CASA \citep{2007ASPC..376..127M}. The contours depict clumps A, B, C, D and E, taken from Figure 1 of S15. We do not use clumps B, C, D and E in this work because they do not allow for accurate estimates of kinematics and volume densities from the plane-of-the-sky projection.}
\label{fig:band7_figure}
\end{figure}

S15 identified 5 molecular clumps from the continuum emission maps where intense star formation is taking place (see Figure 1 in their paper), using a signal to $local$ noise (SNR) cutoff at $5\sigma$. Of these clumps, only clumps A and B have sufficient resolution (number of pixels) to perform the kinematic analysis to estimate the turbulent velocity dispersion (see Figure \ref{fig:band7_figure}). The horizontally elongated structure of clump B rules out the possibility of a spherical approximation to its volume which we otherwise cannot estimate, not to mention that its location does not correlate well with the CO$\,$(5-4) flux map. $\Clump$, on the other hand, has a strong correlation with CO$\,$(5-4) flux and appears symmetric, as seen in Figure \ref{fig:band7_figure}. Hence, we restrict our analysis to $\clump$ in this work. We notice that $\clump$ likely coincides with the center of the galaxy and might be its nucleus/forming core. Thus, it might be significantly different in origin from clumps residing in the outer regions of the disk. 

\section{Measurement of the Star Formation Rate}
\label{s:dustsed}
We estimate the SFR surface density ($\sigmasfr$) in $\clump$ by fitting a modified blackbody (MBB) spectrum to the continuum emission from dust in the three ALMA bands. The modified blackbody spectral law can be written as:

\begin{equation}
S_\nu = N_\nu\,\left(\frac{2h\nu^3}{c^2}\right)\,\frac{1}{e^{h\nu/k_{\mathrm{B}}T}-1}\,\nu^{\emiss}\,,
\label{eq:sed}
\end{equation}

where $\nu$ is the rest-frame frequency, $S_\nu$ is the flux density of the clump, $N_\nu$ is the normalization parameter (includes dust opacity), $T$ is the dust temperature and $\emiss$ is the emissivity index ($\emiss = 0$ corresponds to a blackbody) \citep{1984ApJ...285...89D,2010A&A...523A..78D}. Since $\clump$ lies near the center of the galaxy where the background contribution may be high, we subtract the underlying (disc) CO emission from $\clump$. For background subtraction, we mask the $\clump$ and then smooth the image by convolving it with a large Gaussian kernel. Then, we subtract the smoothed image from the original image to get the background subtracted image. In order to ensure that we do not over or under-subtract, we reiterate this procedure multiple times with different kernels. 

The flux density can be integrated over the whole infrared (IR) range (8--1000 $\mu$m) to get dust luminosity \citep{1956AJ.....61...97H,1968ApJ...154...21O,2002astro.ph.10394H}:

\begin{equation}
\label{eq:lumi}
L_\mathrm{FIR} = \frac{4\pi{D_\mathrm{L}}^2}{1+z} \int_{8\,\micron}^{1000\,\micron} S_\nu\,d\nu\,,
\end{equation}

where $L_\mathrm{FIR}$ is the far infrared luminosity of the clump, $D_\mathrm{L}$ is the luminosity distance to SDP 81 and $S_\nu$ is the flux density of the clump. However, the available ALMA observations are insufficient to simultaneously constrain the dust parameters -- $T$, $\beta_{\mathrm{em}}$, and $L_{\mathrm{FIR}}$. Therefore, we fix $\emiss = 1.5, 2.0$ and $2.5$, which are the typically used values for starburst galaxies \citep{1983QJRAS..24..267H,2003MNRAS.338..733B,2012MNRAS.425.3094C,2013MNRAS.436.2435S}. We also include observed fluxes at various other wavelengths in the infrared regime, reported in Table 2 of \cite{2014MNRAS.440.1999N}, which covers a longer wavelength baseline ($\lambda_{\mathrm{obs}} = 350 - 2000\,\mathrm{mm}$). Then, using a two-step fitting process: 1.) we fit the galaxy-integrated fluxes to constrain $T$ and $\emiss$; 2) we then adopt the 'galaxy-wide' $T$ and $\emiss$ to fit $\clump$ and determine its far infrared luminosity. The conditions of individual clumps might be very different as compared to the whole galaxy. We lack the spatial resolution for a proper decomposition of the various clumps and assume that the clump conditions (of $\clump$) are identical to the galaxy-wide properties, while being aware that this might not be the case. We include the systematic uncertainty arising from this assumption in our calculation of the far infrared luminosity.

\begin{figure}
\includegraphics[width=1.0\linewidth]{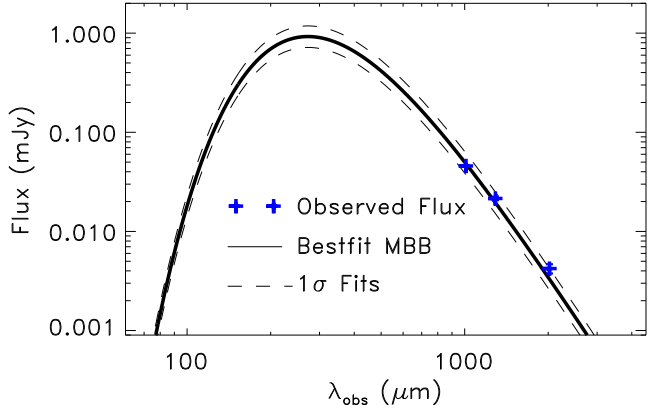}
\caption{SED fit applied to $\clump$ with $T = 39\,\mathrm{K}, \emiss=2.5$ (parameters obtained from SED fit of data from SPIRE \citep{2010A&A...518L...3G}, SMA \citep{2013ApJ...779...25B}, MAMBO \citep{2010Sci...330..800N} and ALMA \citep{2015ApJ...808L...4A}). Observed fluxes for the three ALMA bands are shown in blue. Dashed lines indicate the $1\,\sigma$ uncertainty range of the fit.}
\label{fig:clumpfitbdaries_figure}
\end{figure}

We derive a best-fit temperature $T = 39\pm2\,\mathrm{K}$ for $\emiss = 2.5$, where we use Monte Carlo (MC) simulations (by modeling the uncertainties in the observed fluxes according to a Gaussian distribution) to find the bestfit MBB \citep{OGILVIE1984205,2013AAS...22143103J}. The uncertainty on $T$ arises from the inclusion of the flux at 350 $\micron$ (from SPIRE observations, \citealt{2010A&A...518L...3G}) which falls partially under the cold temperature dominated regime (see Figure 6 of \citealt{2015MNRAS.452.2258D}). Figure \ref{fig:clumpfitbdaries_figure} shows the best-fit MBB we find for $\clump$ from the fits, along with the $1\,\sigma$ uncertainty range we derive from the $1\,\sigma$ uncertainty of the far infrared luminosity ($L_{\mathrm{FIR}}$) obtained using MC error propagation. 

We obtain nearly identical values of ($T, \emiss$) from the two fitting methods we use: pixel-by-pixel and whole clump. In the pixel-by-pixel algorithm, we find the SFR in each pixel by fitting the MBB using the best-fit ($T, \emiss$) values. Then, we sum the SFRs from each pixel to get SFR for the whole clump. On the contrary, in the whole clump fitting algorithm, we first sum the fluxes from each pixel and then find the SFR by fitting the MBB using the best-fit ($T, \emiss$) values. The SFRs we obtain from both the methods agree well with each other, within $\pm\,3$\%.

Using equation \ref{eq:lumi}, we derive $L_{\mathrm{FIR}} = \big(2.25^{+0.95}_{-0.47}\big)\times10^{11}\,\lsol$ for $\clump$, where the uncertainty includes statistical as well as systematic errors. The observed SFR surface density we find using the $L_{\mathrm{FIR}}-\mathrm{SFR}$ relation from \cite{1998ARA&A..36..189K} is $\sigmasfr = 555^{+197}_{-120}\,\sfrunits$. Since this relation used a Salpeter IMF \citep{1955ApJ...121..161S}, we adjust its coefficient by a factor of 1.6 downward to adopt to "the Chabrier IMF" \citep{2003ApJ...586L.133C} to be consistent throughout our work \citep{2010MNRAS.403.1894D,2010A&A...523A..78D}. The resulting SFR surface density we get is $\sigmasfr = 357^{+135}_{-85}\,\sfrunits$. These values are representative of intense star formation and are expected for the central regions of high-redshift starburst galaxies \citep{2017MNRAS.469.4683C,2017A&A...604A.117C}.

To reinforce our estimation of flux densities in the three ALMA bands as obtained after background subtraction, we also model the fluxes using an n-S\'ersic profile ($R^{1/n}$) for the disc, with a Gaussian added to it for $\clump$ \citep{1968adga.book.....S,1993MNRAS.265.1013C,1999A&A...352..447C,2001MNRAS.326..869T,2006MNRAS.373..632A}. The S\'ersic index ($n$) we obtain for $\lambda_{\mathrm{obs}} = 1.0\,\mathrm{mm}$ continuum is $n\,\sim0.5$. Although our result is lower than the median value reported in \citealt{2016ApJ...833..103H} ($n\approx1\pm0.2$, see also \citealt{2018MNRAS.tmp..306P}), it is consistent with the S\'ersic indices found in several high-redshift galaxies (see Table 1 of \citealt{2016ApJ...833..103H}). Through this composite profile, the fluxes we obtain for $\clump$ for the three bands are similar to those obtained through background subtraction discussed above, within $\pm12\%$. Since this difference in flux densities is negligible, the resulting $L_\mathrm{FIR}$ and $\Sigma_{\mathrm{SFR}}$ from this method are similar to those quoted above, within the uncertainties.

\subsection{Gas Mass and Clump Size from Continuum Emission}
\label{s:magdis_and_galfit}
Apart from the SFR surface density, one can also estimate the gas mass ($M_{\mathrm{gas}}$) and size of the clump ($R$) using the SED fits and continuum maps, respectively. Since we have an excellent coverage of the Rayleigh-Jeans regime, we use our best-fit MBB to estimate the dust mass of $\clump$ by using equation 6 of \cite{2012ApJ...760....6M} and appropriate rest-frame dust mass absorption coefficients for the three ALMA bands, from Table 6 of \cite{2001ApJ...554..778L}. Then, we use a typical gas-to-dust conversion ratio of 150 to get the gas mass in this clump \citep{2000MNRAS.315..115D,2015MNRAS.452.2258D,2017A&A...608A..15B}. For the three ALMA bands, the gas masses we thereby obtain are $M_{\mathrm{gas}} = (2.3-4.9)\times10^8\,\msol$. This is consistent with the value of $M_{\mathrm{gas}}$ we obtain from CO$\,$(5-4) (as we discuss in detail in Section \ref{s:gasmass}).

We use the composite disc profile (n-S\'ersic disc + Gaussian clump, as discussed in Section \ref{s:dustsed}) to find an estimate of the size of $\clump$. Since the clump is defined using the $\lambda_{\mathrm{obs}} = 1.0\,\mathrm{mm}$ continuum obtained from ALMA, we use the best-fit composite disc profile at this wavelength and find the size of this clump by assuming that its diameter is equal to the full width at half maximum of the composite profile ($\mathrm{i.e.,\, FWHM} = 2\,R$). Correspondingly, we obtain $R\,\sim0.16\,\mathrm{kpc}$ for $\clump$. This is in good agreement with the size of $\clump$ we find in Section \ref{s:gasmass} by summing up the pixels belonging to $\clump$, as we report in Table \ref{tab:table2}.

\section{Mach Number Estimation}
\label{s:kinematics}
Supersonic turbulence is a key ingredient to star formation because it can compress interstellar gas which leads to the formation of dense cores. On the other hand, it can suppress the global collapse of the clouds, thus significantly reducing the SFR \citep{2004ARA&A..42..211E,2004RvMP...76..125M,2007ARA&A..45..565M,2012A&ARv..20...55H}. The root mean square (RMS) sonic Mach number associated with turbulence in star-forming regions is given by:

\begin{equation}
\mach = \frac{\sigma_\textrm{v,turb}}{c_\textrm{s}}\,,
\label{eq:machnum}
\end{equation}

where $\sigma_{\mathrm{v,turb}}$ is the turbulent velocity dispersion and $c_\mathrm{s}$ is the sound speed. $c_\textrm{s} \propto \sqrt{T}$, where $T$ is the gas temperature. It is difficult to estimate the gas temperature with the current data, however, we can assume it to be between 10--100 K. This assumption is valid for gas temperatures in dense molecular clouds \citep{2004ApJ...606..271G,2005ARA&A..43..677S,2005ApJ...635L.173W,2014ApJ...786..116B,2016A&A...595A..94I,2017ApJ...850...77K}. Using the relation for isothermal sound speed from \cite{2016ApJ...832..143F} for a mean molecular weight of 2.33 \citep{2008A&A...487..993K} and $T_{\textrm{gas}} \approx 10\,\mathrm{K}$, the sound speed is $\sim0.2\,\mathrm{km\,s}^{-1}$ whereas it is  $\sim0.6\,\mathrm{km\,s}^{-1}$ for $T_{\textrm{gas}} \approx 100\,\mathrm{K}$; so we assume the sound speed to be in the range $0.2-0.6\,\mathrm{km\,s}^{-1}$. 

\begin{figure*}
\includegraphics[width=.3\textwidth]{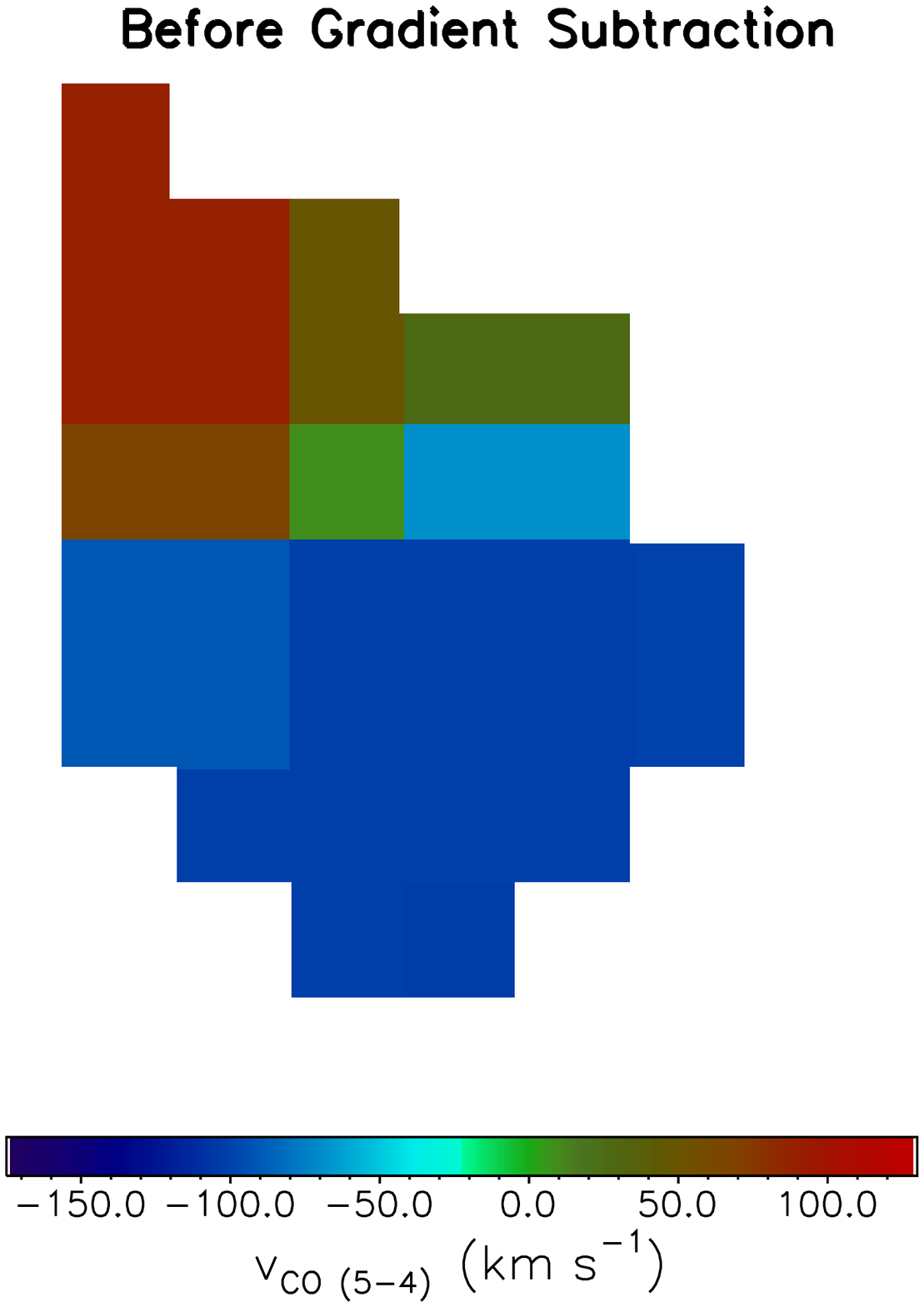}\hfill
\includegraphics[width=.3\textwidth]{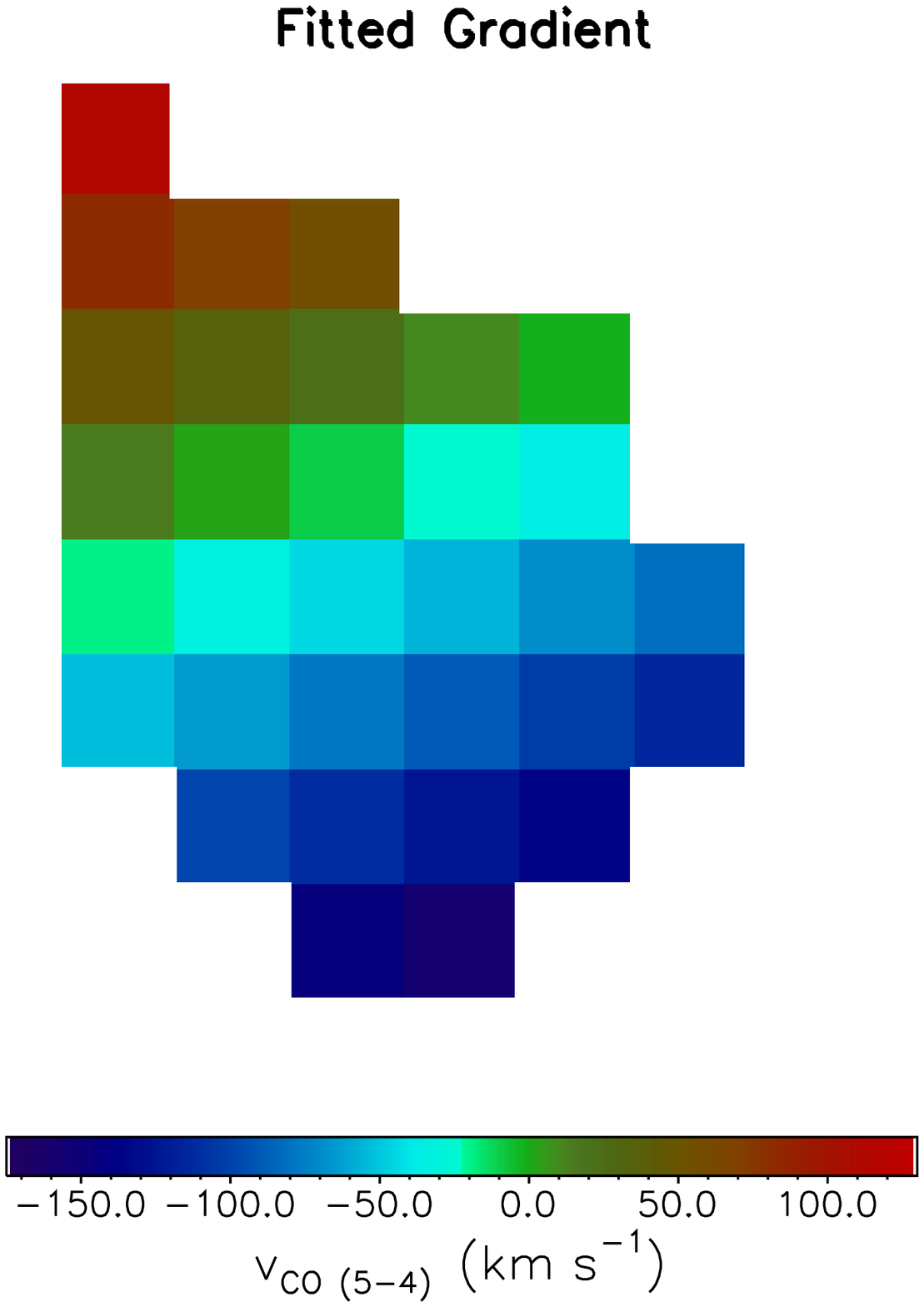}
\hfill
\includegraphics[width=.3\textwidth]{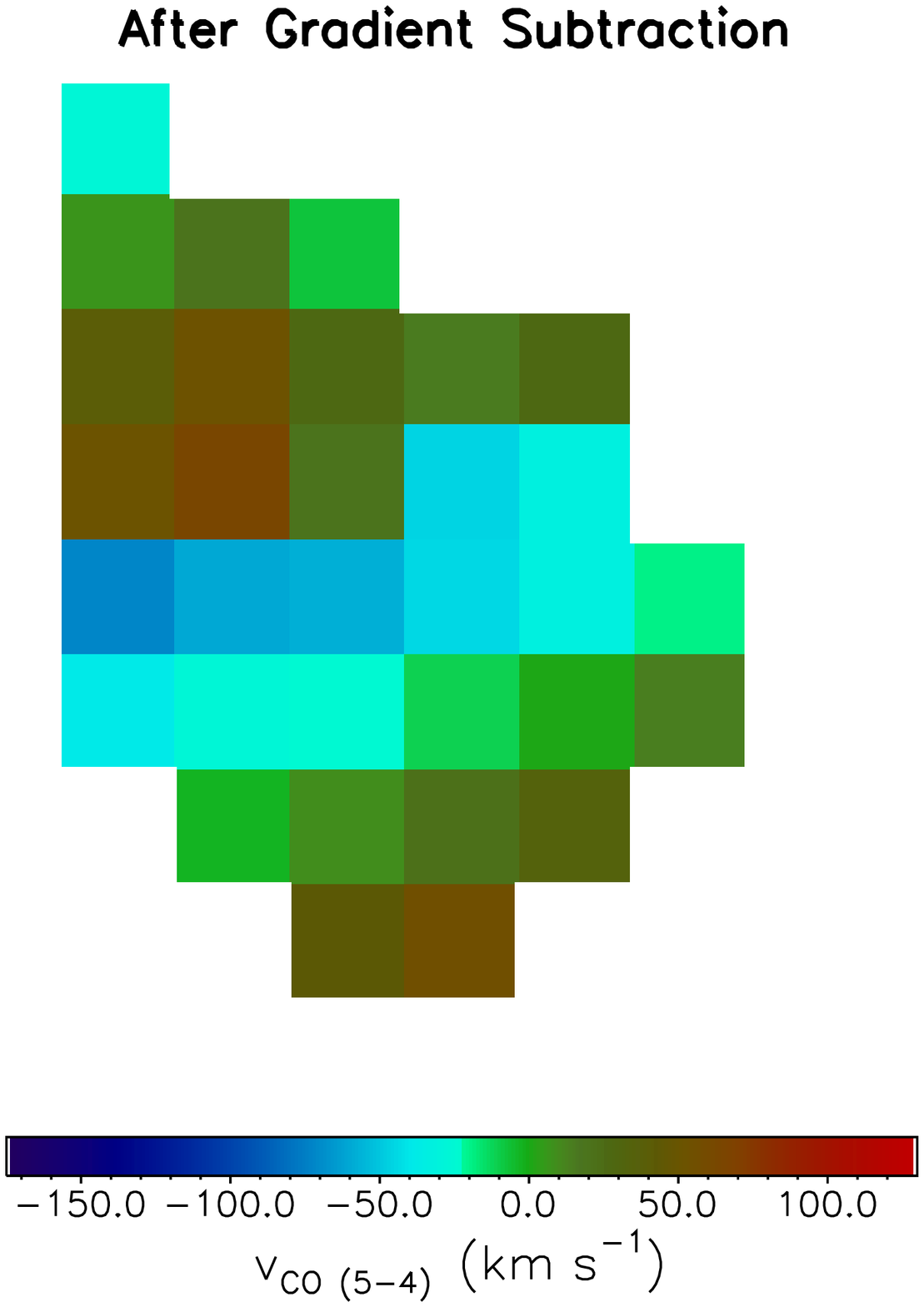}
\caption{\textit{Left Panel:} Centroid CO$\,$(5-4) velocity (in $\mathrm{km\,s}^{-1}$) in the pixels of $\clump$ before gradient subtraction. The size of each pixel is $\sim 0.05\,\kpc$. \textit{Middle Panel:} Fitted large-scale gradient to the clump. \textit{Right Panel:} Velocity of CO$\,$(5-4) in the clump after gradient subtraction. The last panel isolates the turbulent velocities in the plane of the sky. The turbulent velocity dispersion obtained for $\clump$ is $\sigma_{\mathrm{v,turb}} = 37\pm5\,\mathrm{km}\,\mathrm{s}^{-1}$, using which the Mach number calculated is $\mach = \sigma_{\mathrm{v,turb}}/c_{\mathrm{s}} = 96\pm28$.}
\label{fig:grad_figure3}
\end{figure*}

The CO$\,$(5-4) velocity map after source plane reconstruction shows a clear, large-scale gradient running diagonally, as we show in Figure \ref{fig:grad_figure3} (also, Figure 1 of S15). This systematic gradient can be associated with the rotational or shear motion of the gas. To extract the small-scale turbulent features in this clump, we fit a large-scale gradient to the clump and subtract it; similar to the analysis of turbulent velocity dispersion done on the central molecular zone (CMZ) cloud Brick by \cite{2016ApJ...832..143F}. For this purpose, we use the \texttt{PLANEFIT} routine in IDL which performs a least-squares fit of a plane to set of ($x,y,z$) points. In this case, this set is a position-position-velocity (PPV) cube with $x$ and $y$ being the position coordinates of pixels forming $\clump$, and $z$ being the CO$\,$(5-4) velocity of each pixel. We use the standard deviation of residuals after gradient subtraction as the turbulent velocity dispersion:

\begin{equation}
\label{eq:vturb}
\sigma_{\mathrm{v,turb}} = \sqrt{\frac{1}{N-1}\sum_{i = 1}^{N}\Bigg(\big(\mathrm{v}_{\mathrm{bgs}}-\mathrm{v}_{\mathrm{fg}}\big)-\mu\Bigg)^2}\,,
\end{equation}

where $N$ is the number of pixels or resolution elements, $\mathrm{v}_{\mathrm{bgs}}$ is the pixel velocity before gradient subtraction, $\mathrm{v}_{\mathrm{fg}}$ is the velocity of the fitted gradient and $\mu$ is the mean of residuals after gradient subtraction (\textit{i.e.,} $\mu = <(\mathrm{v}_{\mathrm{bgs}}-\mathrm{v}_{\mathrm{fg}})>$). 

\begin{figure*}
\includegraphics[width=1.0\linewidth]{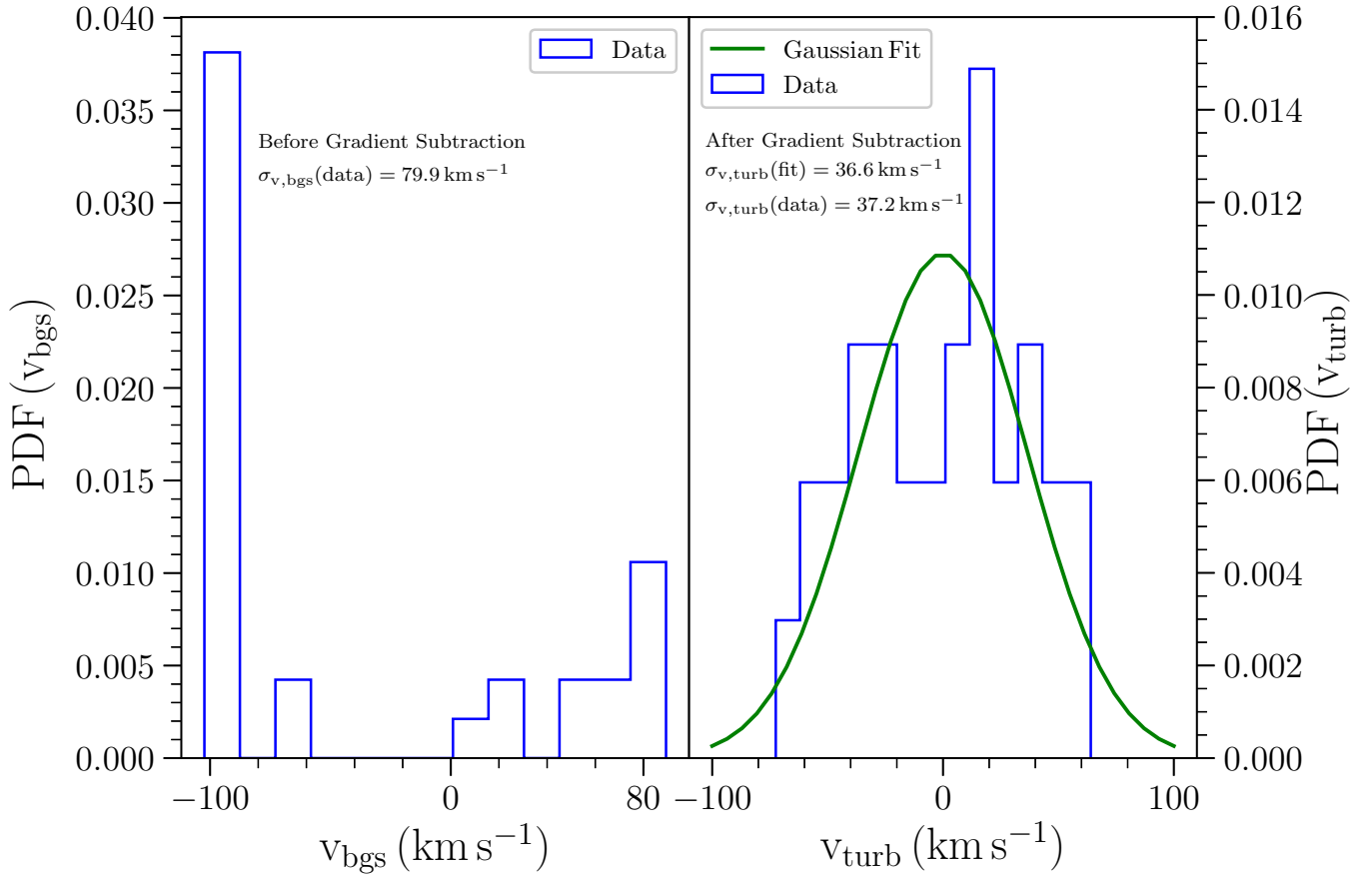}
\caption{PDFs of CO$\,$(5-4) velocities of the pixels in $\clump$ before and after gradient subtraction. The distribution of velocities before gradient subtraction ($\mathrm{v_{bgs}}$) is bimodal; that of velocities after gradient subtraction ($\mathrm{v_{turb}}$) resembles a Gaussian distribution, however, much information cannot be ascertained due to low number of statistics. The width of the Gaussian we fit (right panel) and the data are similar, thus reinforcing the estimate of turbulent velocity dispersion $\sigma_{\mathrm{v,turb}} = 37\pm5\,\mathrm{km\,s^{-1}}$.}
\label{fig:pdf_turbulentvelocities}
\end{figure*}

The turbulent velocity before subtracting the gradient is $\sigma_{\mathrm{v,bgs}}=80\pm10\,\mathrm{km\,s}^{-1}$. From the gradient subtraction algorithm, we obtain the turbulent velocity dispersion in $\clump$ as $\sigma_{\mathrm{v,turb}} = 37\pm5\,\mathrm{km\,s}^{-1}$, where the $1\,\sigma$ error is the standard deviation calculated using \citep{lehmann1998}:

\begin{equation}
\textrm{s}_{\sigma_{\textrm{v}}} = \sigma_\mathrm{v} \cdot \frac{\Gamma\big( \frac{N-1}{2} \big)}{ \Gamma(N/2) } \cdot \sqrt{\frac{N-1}{2} - \left( \frac{ \Gamma(N/2) }{ \Gamma\big( \frac{N-1}{2} \big) } \right)^2 }\,,
\label{eq:sdofsd}
\end{equation}

where $\Gamma(N)$ is the Gamma function. We also find the uncertainty on $\sigma_{\mathrm{v,turb}}$ through MC simulations and note that the result is consistent with the value we obtain from the analytical equation, within $\pm8\,$\%. Figure \ref{fig:grad_figure3} shows the velocity field across $\clump$ before gradient subtraction, fitted gradient velocities and velocity field after gradient subtraction. By construction, the residuals after gradient subtraction are evenly spread around $0$, as is also clear from the PDF of $\sigma_{\mathrm{v,turb}}$ plotted in Figure \ref{fig:pdf_turbulentvelocities}. From Figure \ref{fig:pdf_turbulentvelocities}, we note that the distribution of velocities (in the pixels of $\clump$) before gradient subtraction is highly non-Gaussian and bimodal, while that of the velocities after gradient subtraction is more consistent with a Gaussian distribution. However, due to low-number statistics, it is hard to infer much information from this distribution; some non-Gaussian contributions may still remain after gradient subtraction, because it only removes the largest-scale mode of systematic shear or rotation. Nonetheless, this distribution is in agreement with velocity PDFs obtained for simulations of supersonic turbulence which are also Gaussian in nature \citep{2000ApJ...535..869K,2013MNRAS.436.3167F} and the non-Gaussian components can arise from small-scale rotational or shear modes, or due to the intrinsic features of turbulence (see Section 3.2.2 of \cite{2016ApJ...832..143F} and references therein). Additionally, we note that the width of the Gaussian we fit for the PDF of velocities after gradient subtraction matches well with what we find using the data ($\mathrm{i.e.,}\,\sigma_{\mathrm{v,turb}} (\mathrm{fit}) \approx \sigma_{\mathrm{v,turb}} (\mathrm{data})$).

We also note that the turbulent velocity dispersion we calculate is in agreement with the velocity dispersion of $30\pm9\,\mathrm{km\,s}^{-1}$ calculated by S15 for this clump using the 2$^{\mathrm{nd}}$ moment map ($\mathrm{i.e.,}$ the dispersion along the line of sight, after correction for beam smearing). The velocity dispersion we calculate is in the plane of the sky. A consensus between velocity dispersions using the two methods imply that $\clump$ can be considered isotropic and it is fair to approximate it as a sphere.

Using this turbulent velocity dispersion, we obtain a turbulent Mach number $\mach = 96\pm28$. Although this is quite high compared to nearby galaxies (see \cite{2012ARA&A..50..531K} and references therein), it falls in the range of Mach numbers associated with starburst galaxies \citep{2004ApJ...606..271G,2007ApJ...671..303B,2009ApJ...697..115C,2009ApJ...706.1364F,2010Natur.463..781T}. Given the high redshift of SDP 81 and previous works highlighting intense star formation, it is not unusual to obtain Mach numbers near 100. In fact, it implies that the role of turbulence becomes more important at the epoch near the maximum star formation in the history of the Universe \citep{2003MNRAS.339..312S,2014ARA&A..52..415M,2017Natur.548..430F}.

\subsection{Resolution Check for Gradient Fit and Subtraction Algorithm}
The gradient fit and subtraction algorithm works accurately only if the resolution is sufficient. For the CMZ cloud Brick, the resolution scale was in sub-parsecs \citep{2016ApJ...832..143F} whereas it is in sub-kiloparsecs for SDP 81. We check if the low number of resolution elements in our data affects our measurements, since the velocity dispersion calculated might vary by more than 20\% if sufficient number of pixels are not available to resolve the clump. To investigate whether we have enough pixels to be operating in the saturated regime (where velocity dispersion does not change by more than 20\% when the number of resolution elements are altered), we perform a resolution degradation on $\clump$ by creating artificial 'superpixels' (merging nearby pixels to make a bigger pixel) and then applying the gradient fit subtraction algorithm. 

\begin{figure}
\includegraphics[width=1.0\linewidth]{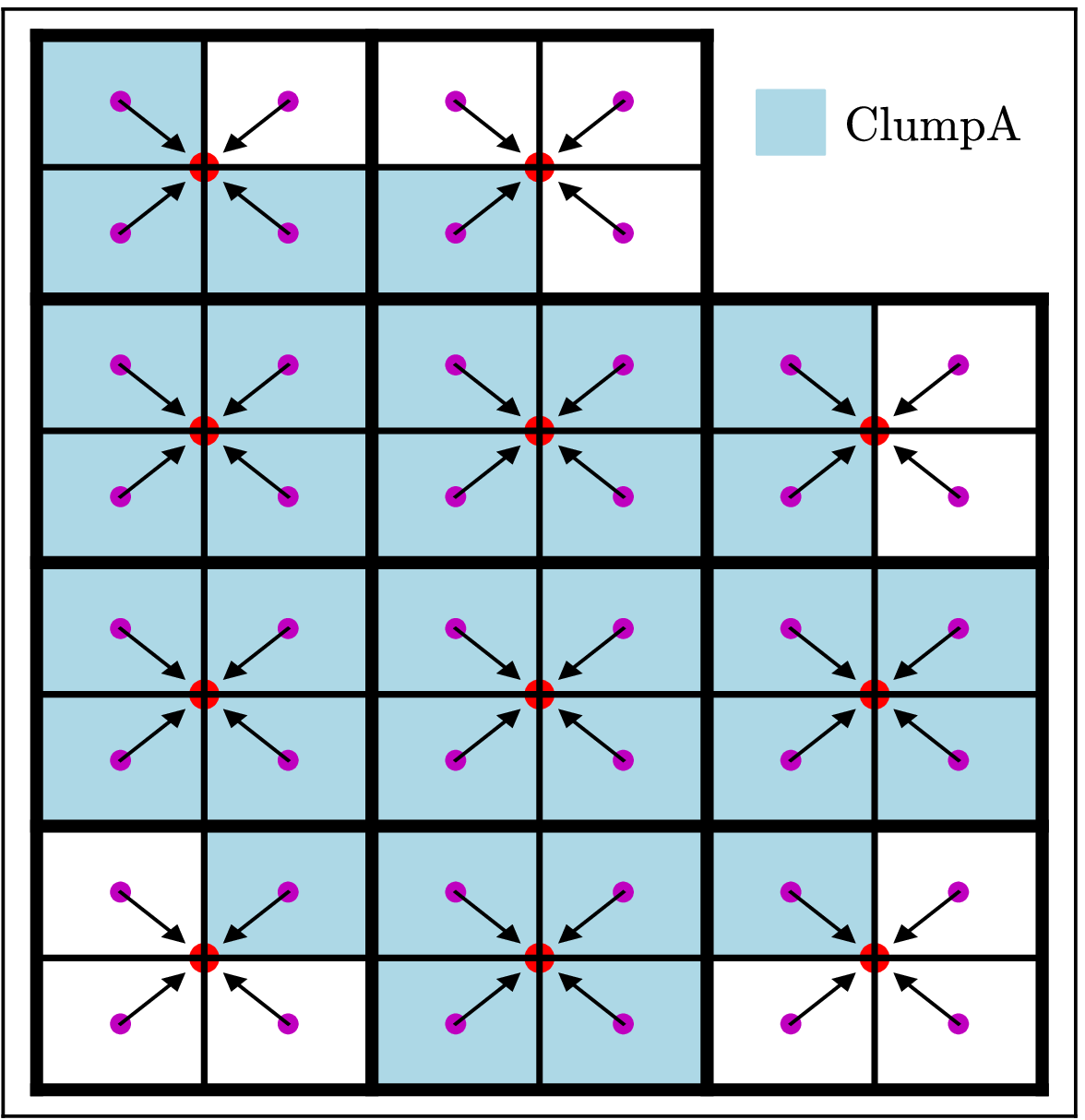}
\caption{First resolution degradation through creation of 11 superpixels, by merging nearby pixels using a flux-weighted averaging method (see equation \ref{eq:fluxmerge}). Thick lines denote the boundaries of superpixels whereas thinner lines denote the boundaries of pixels. Red dots depict the center of superpixels and purple dots depict the center of pixels. Pixels belonging to $\clump$ are shown in blue; they are the same as those shown in Figure \ref{fig:grad_figure3}. Pixels outside the clump are shown in white. Arrows show the movement of a pixel when it is merged with other pixels to create a superpixel.}
\label{fig:gradientmerge_superpixel}
\end{figure}

For the first degradation ($1/4$ resolution), we merge four nearby pixels into one (making a square shape, see Figure \ref{fig:gradientmerge_superpixel}). While the center of a superpixel is the centroid of the four constituent pixels, its  CO$\,$(5-4) velocity is the flux-weighted average of CO$\,$(5-4) velocities in the constituent pixels:

\begin{equation}
\label{eq:fluxmerge}
v_{\mathrm{spix}} = \Bigg(\frac{\sum_{i = 1}^{4} (S_i\,v_i)}{\sum_{i=1}^{4} S_i}\Bigg) \,,
\end{equation}

where $v_{\mathrm{spix}}$ is the velocity of the superpixel, $S_i$ is the flux of the $i^{\mathrm{th}}$ constituent pixel and $v_i$ is its velocity. At places where pixels belonging to the $\clump$ cannot make a square by themselves, we use pixels from outside the clump, but mask their flux to be 0. Thus, for such constituent pixels, $S_i = 0$, so these pixels do not contribute to the sums in equation \ref{eq:fluxmerge}.

\begin{figure}
\includegraphics[width=1.0\linewidth]{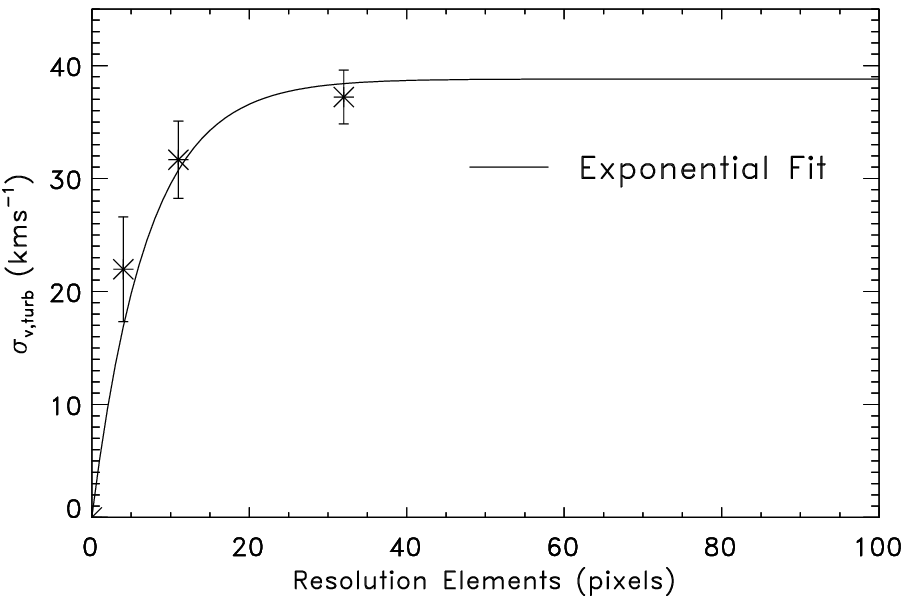}
\caption{Turbulent velocity dispersion at different resolutions, fitted with a growing exponential function of the form $A_0\,(1-\textrm{e}^{-N/N_0})$, where $N$ is the number of resolution elements. $N_0\,\sim\,7$ for the best fit model, which implies the velocity dispersion does not change by more than 5\% for $N\,\gtrsim\,20$.}
\label{fig:resolution_figure}
\end{figure}

We do this process twice: decreasing the resolution to (1/4) and (1/16) of the original resolution. This results in a total of 11 and 4 superpixels after the first and second resolution degradation, respectively. Figure \ref{fig:resolution_figure} shows the turbulent velocity dispersions we get at the three resolutions. We fit a growing exponential of the form $A_\mathrm{0}(1-e^{-N/N_0})$ (where, $N$ is the number of resolution elements) to this data. Since decreasing the resolution by $(1/4)$ does not alter the velocity dispersion by $\gtrsim\,20$\% (as we notice from Figure \ref{fig:resolution_figure}), we confirm that we have enough pixels to resolve this clump with acceptable accuracy. 

\section{Gas Mass and Freefall Time From CO (5-4)}
\label{s:gasmass}
The total gas mass is an essential parameter which goes in all the star formation relations we test in a later section (see Section \ref{s:compare}). It can be estimated by following the CO$\,$(1-0) emission in the star-forming region \citep{2002ApJ...569..605C,2013ApJ...779...43P,2014ApJ...785...44M,2017ApJ...836...66S}. From Figure 1 in \cite{2015MNRAS.453L..26R}, and that in \cite{2015MNRAS.452.2258D}, we notice that there is a significant presence of CO$\,$(1-0) emission at the position of $\clump$. However, the CO$\,$(1-0) data was obtained by the Karl G. Jansky Very Large Array (VLA) at a lower resolution than ALMA \citep{2011MNRAS.415.3473V} and cannot be used for kinematic analysis. Thus, we rely on ALMA observations of CO$\,$(5-4) transition (observed at a frequency of $142.57\,\mathrm{GHz}$ in ALMA Band 4), to estimate the gas mass of $\clump$. It should be noted that CO$\,$(5-4) is generally a poor tracer of the total diffuse molecular gas, but is bright and easily observable at high redshift \citep{2015A&A...577A..46D,2015ApJ...802L..11L,2017A&A...608A.144Y}.

We follow the \cite{1992Natur.356..318S,1992ApJ...398L..29S} relation between line luminosity and integrated flux density of CO$\,$(5-4):

\begin{equation}
\label{eq:solomonequation}
\lcob = 3.25\times10^7\,S_{v}\,\Delta\,v\,\frac{{D_{\mathrm{L}}}^2}{{\nu_{\mathrm{obs}}^2}{(1+z)}^3}\,,
\end{equation}

where $\lcob$ is the line luminosity in $\mathrm{K}\,\mathrm{km\,s}^{-1}\,\pc^2$, $S_v\Delta v$ is the velocity integrated flux density of CO$\,$(5-4) after subtraction of background emission, in $\mathrm{Jy}\,\mathrm{km\,s}^{-1}$, $D_\mathrm{L}$ is the luminosity distance in $\mathrm{Mpc}$ and $\nu_{\mathrm{obs}}$ is the observed frequency of transition in $\mathrm{GHz}$. The line luminosity we obtain is $\lcob = (5.04\pm1.10)\times10^8\,\mathrm{K}\,\mathrm{km\,s}^{-1}\,\pc^2$. Since the transition we observe with ALMA at $z\,\approx$ 3 is higher than the ground (1-0) transition, we introduce an appropriate line ratio factor (defined as the ratio of line luminosity of CO$\,$(5-4) to that of CO$\,$(1-0)), $r_{_{54}} = 0.28\pm0.05$. This value was derived for $\clump$ by S15, where the authors use velocity and magnification maps from the lens model prepared by \cite{2015MNRAS.452.2258D}. This value falls in the typical range of values of $r_{_{54}}$ for SMGs (see \cite{2013ARA&A..51..105C} and references therein).

To get the gas mass from the line luminosity, we use an appropriate CO to H$_2$ conversion factor $\alpha_{\textrm{CO}}$. Although there is a high uncertainty in the value of this factor for nearby as well as high-redshift galaxies \citep{2012MNRAS.426.2601P,2012MNRAS.421.3127N}, the suggested values based on observations of SMGs lie in the range $\sim\,0.8$--$1.0\,\msol$ per ($\mathrm{K}\,\mathrm{km\,s}^{-1}\,\pc^2$) \citep{1998ApJ...507..615D,2005ARA&A..43..677S,2008ApJ...680..246T,2011ApJ...740L..15M,2012ApJ...760...11H,2013ARA&A..51..105C,2013ARA&A..51..207B,2013MNRAS.429.3047B}, which is less by a factor of $\sim4$ than the typical value used for Milky Way clouds and nearby galaxies. \cite{2015MNRAS.452.2258D} used a conversion factor of unity (in the same units) for SDP 81, while \cite{2015PASJ...67...93H} used a value of 0.8. Further, we notice that $\clump$ falls on top of the starburst sequence of the $\siggas-\sigsfr$ relation populated by local ultra-luminous infrared galaxies (ULIRGs) and SMGs \citep{2010ApJ...714L.118D}. This further justifies the choice of $\alpha_{\textrm{CO}}\,\sim\,0.8-1\,\msol$ per $(\mathrm{K\,km}\,\mathrm{s}^{-1} \mathrm{pc}^2)$.

Keeping these studies in mind, we assume $\alpha_{\mathrm{CO}} \approx 0.9\pm0.2\,$ $\msol$ per ($\mathrm{K\,km}\,\mathrm{s}^{-1} \mathrm{pc}^2$), which suggests an H$_2$ mass of $(4.5\pm1.0)\times10^8\,\msol$ \footnote{This gas mass is essentially in agreement as that obtained by S15 for $\clump$. However, due to a typographical error, the gas masses reported in the last column of table 1 of S15 have to be rearranged. The gas masses reported are in the order D-C-A-B-E.}. Accounting for the contribution to the gas by He, we further increase the H$_2$ mass obtained so far by 36\% to get the total gas mass for $\clump$ as $(6.2\pm1.4)\times10^8\,\msol$. This value is in good agreement with the gas mass found out using SED fitting in Section \ref{s:dustsed}. The gas surface density we derive is $\siggas = (8.6\pm1.9)\times10^9\,\msol\,\kpc^{-2}$, where we calculate and sum the area of all pixels which constitute $\clump$ \footnote{The size of 1 pixel is $\sim0.05\,\kpc$. There are 32 pixels in this clump.}. Moreover, the size of $\clump$ we obtain in this manner is $R\,\sim\,0.15\pm0.02\,\mathrm{kpc}$, in excellent agreement with the size we find through composite disc profile fitting in Section \ref{s:magdis_and_galfit}. Assuming $\clump$ to be spherical (see section \ref{s:kinematics} for a discussion on the validity of this assumption), we calculate its density to be $\rho = M_{\mathrm{gas}}/V = (2.9\pm0.6)\times10^{-21}\,\mathrm{g}\,\mathrm{cm}^{-3}$, where $V$ is the volume of the spherical clump. 

To establish whether the cloud could be collapsing, we estimate the virial parameter $\alpha_{\mathrm{vir}}$, which is the ratio of twice the kinetic energy to the gravitational energy \citep{2012ApJ...761..156F}. Using the definition from \cite{1992ApJ...395..140B}, the virial parameter can also be given by: 

\begin{equation}
\label{eq:virial}
\alpha_{\mathrm{vir}} = \frac{5\sigma^{2}_{\mathrm{v,tot}}}{4\pi GR^2\rho}\,,
\end{equation}

where, the velocity dispersion $\sigma_{\mathrm{tot}}$ is the total thermal and turbulent velocity dispersion including the shear component (\textit{i.e.,} turbulent velocity dispersion before gradient subtraction, $\sigma_{\mathrm{v,bgs}}$). However, in this clump, since the turbulent velocity dispersion $\sigma_{\mathrm{v,bgs}} \gg c_\mathrm{s}$, it implies that the total velocity dispersion can be approximated as $\sigma_{\mathrm{v,tot}} \approx \sigma_{\mathrm{v,bgs}}$ \citep{2005ApJ...630..250K,2012ApJ...761..156F}. The virial parameter we thus obtain is $\alpha_{\mathrm{vir}} = 0.63\pm0.13$ < 1, implying the cloud is strongly gravitationally bound and likely undergoing collapse. For such a cloud, the freefall time can be given by \citep{2011ApJ...743L..29H,2013ApJ...770..150H,2014ApJ...796...75C}:

\begin{equation}
t_{\mathrm{ff}} = \sqrt{\frac{3\pi}{32G\rho}}\,,
\label{eq:tff}
\end{equation}

where G is the gravitational constant. From this equation, we obtain a freefall time of $1.3\pm0.1\,\mathrm{Myr}$. This value is in agreement with freefall times calculated for other high $z$ starbursts (see Table 4 of \citealt{2013ApJ...779...89K}).

We summarize all the parameters going into predictions of SFR surface density in various star formation relations in Table \ref{tab:table2}. 

\section{Comparison of observed SFR surface density with theoretical predictions by K98, KDM12 and SFK15}
\label{s:compare}
\begin{figure}
\includegraphics[width=1.0\linewidth]{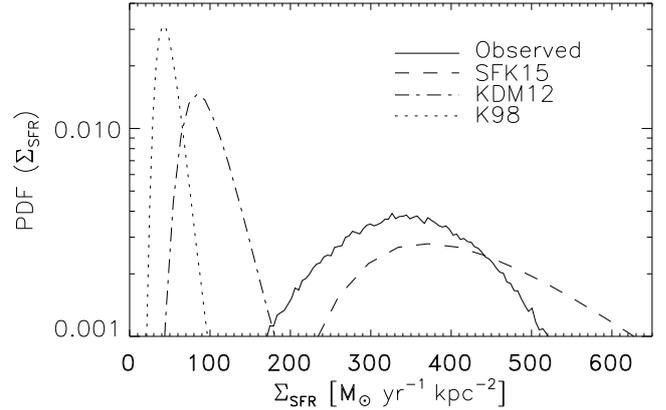}
\caption{Probability distribution function (PDF) of $\sigmasfr$ estimated from dust SED fitting (solid curve) and those predicted from K98 (dotted curve, equation \ref{eq:kslaw}), KDM12 (dot-dashed curve, equation \ref{eq:kdmlaw}) and SFK15 (dashed curve, equation \ref{eq:sfklaw}) relations. The PDFs were calculated using MC simulations with a sample size of 100,000.}
\label{fig:pdf_allsfrs}
\end{figure}

We compare the SFR surface density obtained through dust SED fitting with star formation relations proposed for nearby and high-redshift galaxies in Figure \ref{fig:pdf_allsfrs}. The probability density function (PDF) of the measured SFR surface density in $\clump$ is shown as the solid line, and we compare it with the predictions of SFR surface density by three popular star formation relations in the same plot. These PDFs were calculated using Monte Carlo simulations with a sample size of 100,000 and included systematic errors on the SFR surface densities.

The Kennicutt-Schmidt (KS or K98) relation is given by \citep{1959ApJ...129..243S,1998ApJ...498..541K}:

\begin{equation}
\centering
\label{eq:kslaw}
\Sigma_{\textrm{SFR}} = 2.53\times10^{-4}\, {\Sigma_{\textrm{gas}}}^{1.4\pm0.15}\,,
\end{equation}

The distribution of SFR surface density ($\sigmasfr$) obtained using equation \ref{eq:kslaw} is shown as the dotted line in Figure \ref{fig:pdf_allsfrs}. The mean SFR surface density we calculate from the KS relation is $\sigmasfr = 84\pm27\,\sfrunits$. Since equation \ref{eq:kslaw} is based on the Salpeter IMF, we correct the SFR surface density for a Chabrier IMF (similar to that done in section \ref{s:dustsed}) and obtain $\sigmasfr = 52\pm17\,\sfrunits$.\footnote{This distribution does not take into account the uncertainty on the power law index in equation \ref{eq:kslaw} because we notice that it becomes highly skewed when this uncertainty is randomized. In that case, the $16^{\mathrm{th}}$ and $84^{\mathrm{th}}$ percentile values of $\sigmasfr$ are $1$ and $133\,\sfrunits$, respectively.} We find that the KS relation underestimates SFR surface density by a factor $\gtrsim\,3.3$, with respect to the observed SFR surface density in this clump, even when the $1\,\sigma$ uncertainty is taken into account. Numerous studies discuss the breakdown of the KS relation on the scales of $\sim100\,\pc$ in local \citep{2010ApJ...722L.127O,2011ApJ...733...87S,2014ApJ...786...56B,2015ApJ...799...11X} and high-redshift environments \citep{2007ApJ...671..303B,2010ApJ...714L.118D,2010MNRAS.407.2091G}. 

\begin{table}
\caption{Results for $\Clump$ in SDP 81 (see Figure \ref{fig:band7_figure}) with mean and $1\,\sigma$ errors.}
\label{tab:table2}
\def\arraystretch{1.1}
\setlength{\tabcolsep}{15.0pt}
\begin{tabular}{|lr|}
Parameter & Value\\	
\hline
\hline
$\emiss$ & $2.5$\\
$T$ & $39\pm2\,\mathrm{K}$\\
$L_\mathrm{FIR}$ & $\big(2.3^{+1.0}_{-0.5}\big)\times10^{11}\,\lsol$\\
$\Sigma_{\mathrm{SFR}}\, ({\mathrm{observed}})^{\mathrm{a}}$ & $357^{+135}_{-85}\,\msol\,\yr^{-1}\,\kpc^{-2}$\\
$A$ & $(7.2\pm1.5)\times10^{-2}\,\mathrm{kpc}^2$\\
$R$ & $(1.5\pm0.2)\times10^{-1}\,\mathrm{kpc}$\\
$\sigma_{\textrm{v,bgs}}^{\mathrm{(b)}}$ & $80\pm10\,\mathrm{km\,s}^{-1}$\\
$\sigma_{\textrm{v,turb}}^{\mathrm{(c)}}$ & $37\pm5\,\mathrm{km\,s}^{-1}$\\
$c_{\textrm{s}}$ & $0.4\pm0.2\,\mathrm{km\,s}^{-1}$\\
$\mach$ & $96\pm28$\\
\textit{b}&0.4\\
${S_{\mathrm{v,CO}\,(5-4)}v}^{\mathrm{(d)}}$ & $8.4\pm1.8\, \mathrm{mJy}\,\mathrm{km\,s}^{-1}$\\
$\lcoa$ & $(1.4\pm0.1)\times10^8\,\mathrm{K}\,\mathrm{km\,s}^{-1}\,\pc^2$\\
$r_{54}$ & $0.28\pm0.05$\\
$\lcob$ & $(5.0\pm1.1)\times10^8\,\mathrm{K}\,\mathrm{km\,s}^{-1}\,\pc^2$\\
$\alpha_{_{\mathrm{CO}}}$ & $0.9\pm0.2 \,\msol\,{(\mathrm{K}\,\mathrm{km\,s}^{-1}\,\pc^2)}^{-1}$\\
$M_{\mathrm{gas}}$ & $(6.2\pm1.4)\times10^8\,\msol$\\
$\Sigma_{\textrm{gas}}$ & $(8.6\pm1.9)\times10^9\,\msol\,\mathrm{kpc}^{-2}$\\
$\rho$ & $(2.9\pm0.6)\times10^{-21}\,\mathrm{g}\,\mathrm{cm}^{-3}$\\
$\alpha_{\mathrm{vir}}$&$0.6\pm0.1$\\
$\tff$ & $(1.3\pm0.1)\times10^{6}\,\mathrm{yr}$\\
$\Sigma_{\mathrm{SFR}}\,(\mathrm{K}98)\,^{\mathrm{(a,e)}}$ & $52\pm17 \,\msol \mathrm{yr}^{-1}\, \kpc^{-2}$\\
$\Sigma_{\mathrm{SFR}}\,(\mathrm{KDM}12)\,^{\mathrm{(f)}}$ & $106\pm37 \,\msol \mathrm{yr}^{-1}\, \kpc^{-2}$\\
$\Sigma_{\mathrm{SFR}}\,(\mathrm{SFK}15)\,^{\mathrm{(g)}}$ & $491^{+139}_{-194} \,\msol \mathrm{yr}^{-1}\, \kpc^{-2}$\\
\hline
\end{tabular}\\
$^\mathrm{(a)}$Corrected for Chabrier IMF.\\
$^\mathrm{(b)}$Large-scale velocity dispersion before gradient subtraction.\\
$^\mathrm{(c)}$Turbulent velocity dispersion after gradient subtraction.\\
$^\mathrm{(d)}$Integrated CO$\,$(5-4) flux after background subtraction.\\
$^\mathrm{(e)}$16$^{\mathrm{th}}$ and 84$^{\mathrm{th}}$ percentiles are $37$ and $64\,\sfrunits$.\\
$^\mathrm{(f)}$16$^{\mathrm{th}}$ and 84$^{\mathrm{th}}$ percentiles are $74$ and $132\,\sfrunits$.\\
$^\mathrm{(g)}$ Errors represent the 16$^{\mathrm{th}}$ and 84$^{\mathrm{th}}$ percentiles.\\
\end{table}

Krumholz, Dekel and McKee (KDM12) showed that the SFR does not only depend on gas surface density but also on the depletion time of the gas under collapse. Their single-freefall time model takes the form:

\begin{equation}
\label{eq:kdmlaw}
\Sigma_{\textrm{SFR}} = f_{\textrm{H}_{2}} \epsilon_{\textrm{ff}} \frac{\Sigma_{\textrm{gas}}}{t_{\textrm{ff}}}\,,
\end{equation}

where $f_{\textrm{H}_{2}}$ is the fraction of gas available in molecular form (assumed to be unity), and $\epsilon_{\textrm{ff}}$ is the SFR per freefall time. They found a best fit $\epsilon_{\textrm{ff}} = 0.015$ (see \citealt{2013ApJ...779...89K}). The freefall time they used is the minimum of the Toomre timescale (equation 8 in KDM12) and the giant molecular cloud (GMC) freefall time (equation 4 in KDM12). The SFR surface density suggested from KDM12 is $\sigmasfr = 106\pm37\,\sfrunits$. We plot the PDF of the predicted SFR surface density from KDM12 as the dot-dashed line in Figure \ref{fig:pdf_allsfrs}. 

As can be seen from Figure \ref{fig:pdf_allsfrs}, both KS and KDM12 relations underestimate the observed SFR surface density in $\clump$. In order to match the SFR surface density predicted by the KS or KDM12 relations with that observed, the dust temperature (which goes into the modified blackbody function, see equation \ref{eq:sed}) should be lowered by $\sim8\,\mathrm{K}$, if emissivity is fixed at the best fit emissivity $\emiss = 2.5$. Another way could be to decrease the emissivity to $1.9$ such that the original best fit temperature ($T = 39\,\mathrm{K}$) can provide a reasonable match of observed SFR surface density with that predicted by KS or KDM12 relations. Although these values of ($T, \emiss$) are not favored by the SED fit, they are in the typical range of dust temperature and emissivity found out for other high-redshift galaxies \citep{2013MNRAS.436.2435S,2014ApJ...792...34O,2015ApJ...802L..11L,2017ApJ...842L..16L}. Furthermore, in order to fit the KS relation to the observed SFR surface density in this clump, we find that the power law should be steeper with an exponent of $n\approx 1.6$ in equation \ref{eq:kslaw}). This is consistent with scaling in the KS relation estimated for other SMG galaxies and starbursts \citep{2007ApJ...671..303B,2017MNRAS.468..920K}. Moreover, \cite{2010ApJ...714L.118D} also proposed that the KS relation underpredicts the star formation efficiency in starburst galaxies by a factor of 10. Similarly, an equivalence between observed SFR surface density and that predicted by KDM12 relation can be obtained if $\epsilon_{\textrm{ff}}$ in equation \ref{eq:kdmlaw} is increased to thrice its best fit value. These discrepancies between observed and predicted SFR surface density motivates us to include the role of turbulence in star formation relations, as suggested by SFK15.

\begin{figure*}
\includegraphics[width=1.0\linewidth, angle=0]{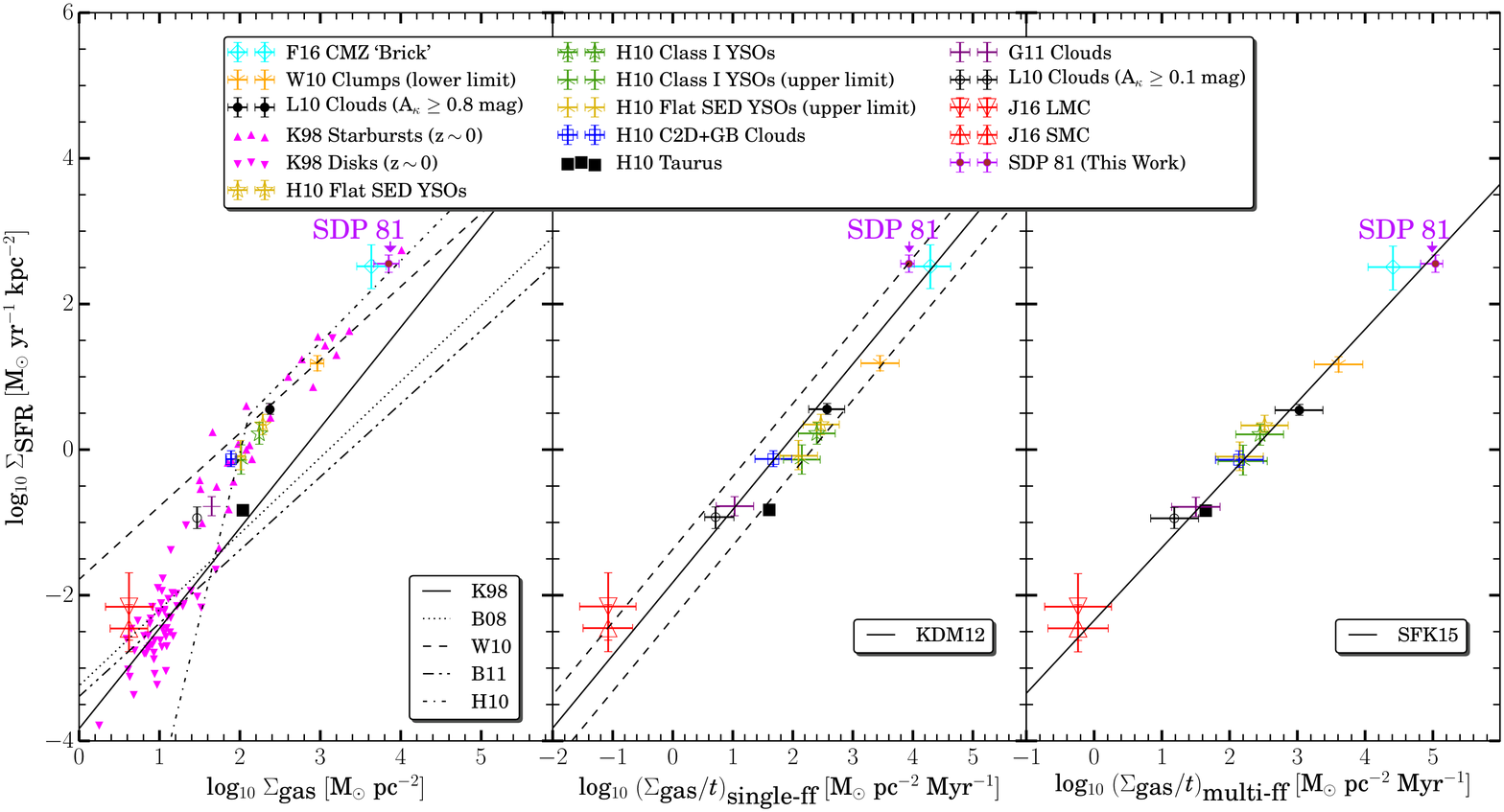}
\caption{\textit{Left-panel:} Observations of $\sigmasfr$ (SFR surface density) in local clouds, plotted against gas surface density. The data is extracted from \protect\citealt{2010ApJ...723.1019H} (H10), \citealt{2010ApJ...724..687L} (L10), \citealt{2010ApJS..188..313W} (W10), \citealt{2011ApJ...739...84G} (G11) and \citealt{2016ApJ...825...12J} (J16). Data for the Brick molecular cloud in the central molecular zone (CMZ) is taken from \protect\citealt{2016ApJ...832..143F} (F16). The K98 (disks and starbursts) data were adjusted in KDM12, \protect\cite{2013ApJ...779...89K} and \protect\cite{2013MNRAS.436.3167F} to the Chabrier IMF, similar to that done for the high-redshift data by \protect\cite{2010ApJ...714L.118D}. The systematic difference between starbursts and the original K98 relation is due to different $\alpha_\mathrm{CO}$ conversion factors used for starbursts and disks (see Section 2 of \protect\citealt{2010ApJ...714L.118D} and Figure 1 of \protect\citealt{2013MNRAS.436.3167F}). SFR relations proposed by \citealt{1998ApJ...498..541K} (K98, corrected for Chabrier IMF), \citealt{2008AJ....136.2846B} (B08), \citealt{2011ApJ...730L..13B} (B11), \citealt{2010ApJS..188..313W} (W10) and \citealt{2010ApJ...723.1019H} (H10) are also shown. \textit{Middle-panel:} Observations of $\sigmasfr$ plotted against the single-freefall time. Solid line depicts the best fit model from KDM12, with $\epsilon_{\textrm{ff}} =  0.015$ (see \protect\cite{2013ApJ...779...89K}); dashed lines illustrate deviations by a factor of 3 from the best fit. \textit{Right-panel:} Observations of $\sigmasfr$ plotted against the turbulence based multi-freefall model proposed by \citealt{2015ApJ...806L..36S} (SFK15). Solid line represents equation \ref{eq:sfklaw}. $\Clump$ analyzed in this work is marked with an arrow in all the three panels. The scatter obtained through a $\chi^2$ minimization routine for K98, KDM12 and SFK15 relations is 50.1, 7.27 and 1.25, respectively.}
\label{fig:all_laws_figure}
\end{figure*}

The SFK15 relation is the result of the combination of gas surface density and density-dependent freefall time determined by KDM12 (see also \citealt{2013ApJ...779...89K}) and the role of turbulence in star-forming regions \citep{2012ApJ...761..156F,2013MNRAS.436.3167F}. SFK15 was able to correlate the scatter present in the KS relation with turbulent motions in gas clouds and using robust fitting techniques, found that $\sigmasfr$ is $\sim0.45\%$ of the multi-freefall gas consumption rate (MGCR) in star-forming regions:

\begin{equation}
\centering
\label{eq:sfklaw}
\Sigma_{\mathrm{SFR}} = \Big(\frac{0.45}{100}\Big)\,\frac{\Sigma_{\mathrm{gas}}}{t_{\mathrm{ff}}} \,\Big({1+b^2\mach^2\frac{\plasmabeta}{\plasmabeta+1}}\Big)^{3/8}\,,
\end{equation}

where $b$ is the turbulent driving parameter ($b = 1/3$ for solenoidal driving and $b=1$ for compressive driving) \citep{2008ApJ...688L..79F,2010A&A...512A..81F,2012ApJ...761..156F,2013MNRAS.436.3167F}. We use a mixed driving mode with $b=0.4$ \citep{2010A&A...512A..81F}. The turbulence term was derived by \cite{2012MNRAS.423.2680M}. In this term, $\plasmabeta$ is the ratio of to thermal to magnetic pressure. It can also be expressed as a ratio of Alfv\'en to sonic Mach numbers: $\plasmabeta = 2\,\mach^{2}_{A}/\mach^2$. The freefall time used in this equation comes from $\mathrm{min}\,(t_{\mathrm{ff,T}}\,, t_{\mathrm{ff,GMC}})$, where $\mathrm{T}$ stands for Toomre and $\mathrm{GMC}$ stands for giant molecular clouds, as used by KDM12. KDM12 showed that the Toomre time is shorter than the GMC freefall time for starburst galaxies. We do not have any estimates of magnetic field strength in this galaxy since it requires polarization or Zeeman measurements of the magnetic field, which are unavailable for SDP 81. For simplicity, we neglect magnetic fields and set $\plasmabeta \to \infty$, leading to $\plasmabeta/(\plasmabeta+1) = 1$. 

The SFK15 relation (equation \ref{eq:sfklaw}) generates a skewed distribution of predicted SFR surface density, as shown in Figure \ref{fig:pdf_allsfrs}. The mean of the distribution is $\sigmasfr = 491^{+139}_{-194}\,\sfrunits$, where the errors represent the $16^{\mathrm{th}}$ and $84^{\mathrm{th}}$ percentiles. As can be seen from Figure \ref{fig:pdf_allsfrs}, the distribution of SFR surface density predicted from SFK15 overlaps to a good extent with the distribution of the observed SFR surface density. The overestimation of SFR surface density by SFK15 can be attributed to ignoring the magnetic field strength, which can reduce the SFR by a factor of $\sim\,2$ \citep{2011ApJ...730...40P,2012ApJ...761..156F,2015MNRAS.450.4035F}. 

Figure \ref{fig:all_laws_figure} depicts the observed SFR surface density ($\sigmasfr$) plotted against gas surface density ($\siggas$) and single and multi-freefall times, overlayed with the three star formation relations, consistent with the Chabrier IMF. We also plot other star formation relations based on gas surface density (\citealt{2008AJ....136.2846B} (B08); \citealt{2010ApJ...723.1019H} (H10); \citealt{2010ApJS..188..313W} (W10) and \citealt{2011ApJ...730L..13B} (B11)) in the first panel of Figure \ref{fig:all_laws_figure}. While the H10 relation can possibly explain the observed SFR surface density in the $\clump$ in SDP 81, it is not universally applicable. Other star formation relations shown in this panel cannot account for observed SFR surface density in all the molecular clouds. Additionally, we also note that SDP 81 lies on the dashed line in the middle panel of Figure \ref{fig:all_laws_figure}, which represent deviations by a factor of 3 from the best fit relation of KDM12 (shown as the solid line). It is evident that there is a large scatter in both the K98 and KDM12 relations, which we calculate from:

\begin{equation}
\chi^2_{\mathrm{red}} = \frac{1}{N_{\mathrm{D}}}\,\sum \Bigg({\frac{\Sigma_{\mathrm{SFR}}\,(\mathrm{observed})-\Sigma_{\mathrm{SFR}}\,(\mathrm{predicted})}{E}}\Bigg)^2\,,
\label{eq:scatter}
\end{equation}

where $E$ is the measured error on $\Sigma_{\mathrm{SFR}}\,(\mathrm{observed})$ and $N_{\mathrm{D}}$ is the number of star-forming regions\footnote{For the calculation of scatter, only those data are included for which $\Sigma_{\mathrm{SFR}}\,(\mathrm{predicted})$ is available for each of K98, KDM12 and SFK15 relations.}. We emphasize that we do not fit any relations to compute the scatter but simply perform a $\chi^2$ minimization routine. From equation \ref{eq:scatter}, we calculate a reduced $\chi^2$ scatter of $\chi^2_{\mathrm{red}} = 50.1$ and $7.27$ for the KS and KDM12 relations, respectively. The final panel in Figure \ref{fig:all_laws_figure} illustrates the turbulence based SFK15 relation. We observe that the characteristics of the $\clump$ in SDP 81 match the SFK15 relation to a good extent. The scatter we obtain for the SFK15 relation is $1.25$. It has significantly reduced as compared to the scatter from KS and KDM12 relations because the SFK15 relation includes systematic variations in the Mach number, as were established by \cite{2013MNRAS.436.3167F}. This highlights the role of turbulence in star-forming regions \citep{2012ApJ...761..156F,2014ApJ...784..112K}. The validity of the multi-freefall star formation relation has been previously supported in an independent work by \cite{2015MNRAS.454.1545B}.

\section{Conclusions}
\label{s:summ}
Using high-resolution (sub-kpc) ALMA data of SDP 81 -- a high-redshift ($z\sim\,3$) lensed galaxy, we have measured the SFR surface density in its biggest and most isotropic clump revealed by the lensing analysis of \cite{2015MNRAS.452.2258D}. Through dust SED fitting by a modified blackbody spectrum of this clump ($\clump$ in S15), we find the best fit dust temperature to be $39\,\mathrm{K}$ when an emissivity index $\emiss = 2.5$ is used. We determine the corresponding SFR surface density of this clump as $\sigmasfr = 357\pm9\,\sfrunits$, which is in the sub-Eddington limit for starburst galaxies at the given redshift. Taking into account the systematic errors resulting from partially cold temperature dominated flux and assuming that this clump has conditions similar to those in the whole galaxy, we obtain $\sigmasfr = 357^{+135}_{-85}\,\sfrunits$. 

Using CO$\,$(5-4) flux and velocity data for this galaxy, we obtain a turbulent velocity dispersion of $\sigma_{\mathrm{v,turb}} = 37\pm5\,\mathrm{km}\,\mathrm{s}^{-1}$, corresponding to a turbulent Mach number $\mach = 96\pm28$. This is somewhat higher than the typical Mach numbers found for local galaxies, but is in agreement with those estimated for high-redshift starbursts. The turbulent velocity dispersion that goes into estimating this Mach number is obtained from large-scale gradient subtraction from the CO$\,$(5-4) velocity, which is in good agreement with the velocity dispersion obtained along the line of sight by S15 after correcting for beam smearing. Using an appropriate CO to H$_2$ conversion factor for this galaxy, we find the gas mass in the clump we study to be $(6.2\pm1.4)\times10^8\,\msol$, which is in agreement with that found out using the best-fit modified blackbody function and the dynamical mass obtained for this clump (S15). 

On testing star-forming relations based on gas mass, freefall times and turbulence (available in literature), we find that the KS relation underpredicts the observed SFR surface density in this clump ($\Sigma_{\mathrm{SFR,KS}} = 52\pm17\,\sfrunits$) by a factor $\gtrsim3.3$, which can be corrected for if the dust temperature is lowered while keeping the emissivity the same or vice-versa. It is also clear that the other star formation relations as plotted in first panel of Figure \ref{fig:all_laws_figure} are not universally applicable. Further, the freefall time based KDM12 relation also underestimates the observed SFR surface density in this clump, giving $\Sigma_{\mathrm{SFR,KDM}} = 106\pm37\,\sfrunits$; however, it can explain the observed SFR if deviations up to a factor of 3 from its best fit model are considered. We also find that the large scatter present in these star formation relations can be explained by turbulence acting in this clump. The turbulence regulated multi-freefall model by SFK15 predicts the SFR surface density as $\Sigma_{\mathrm{SFR,SFK}} = 491^{+139}_{-194}\,\sfrunits$. The overestimation of SFR surface density by SFK15 can be attributed to ignoring magnetic fields while calculating the SFR through equation \ref{eq:sfklaw}. Our findings emphasize the role of turbulence giving rise to the multi-freefall model of the SFR and its consistency with the observed SFR in molecular clouds in local as well as high-redshift galaxies.

\section*{Acknowledgements}
The authors thank the anonymous referee for comments which significantly helped to improve the paper. P.S. acknowledges travel support from the International Programmes and Collaboration Division, BITS Pilani, India\footnote{\url{www.bits-pilani.ac.in/university/ipcd/home}}. C.F. acknowledges funding provided by the Australian Research Council's Discovery Projects (grants DP150104329 and DP170100603), the ANU Futures Scheme, and the Australia-Germany Joint Research Cooperation Scheme (UA-DAAD). E. dC. gratefully acknowledges the Australian Research Council for funding support as the recipient of a Future Fellowship (FT150100079). S.D. is a Rutherford Fellow supported by the UK STFC. The authors also acknowledge the use of WebPlot Digitizer, an extremely useful online image-data mapping tool\footnote{\url{https://automeris.io/WebPlotDigitizer/index.html}}.

This paper uses data from ALMA program ADS/JAO.ALMA \#2011.0.00016.SV. ALMA is a partnership of ESO, NSF (USA), NINS (Japan), NRC (Canada), NSC and ASIAA (Taiwan), and KASI (Republic of Korea) and the Republic of Chile. The JAO is operated by ESO, AUI/NRAO and NAOJ.

\bibliographystyle{mnras}
\bibliography{references}
\label{lastpage}
\end{document}